\documentclass[twocolumn, aps, prc, superscriptaddress,floatfix]{revtex4-1}

\usepackage{multirow}
\usepackage{epsfig}
\usepackage{amsmath}

\usepackage{latexsym}
\usepackage{feynmp}

\usepackage[normalem]{ulem}  
\usepackage[dvips]{color} 

\renewcommand\sout{\bgroup \color{red} \ULdepth=-.5ex \ULset}

\begin{document}

\title{Masses of the doubly heavy tetraquarks in a constituent quark model}

\author{Woosung Park}
\email{diracdelta@yonsei.ac.kr}\affiliation{Department of Physics and Institute of Physics and Applied Physics, Yonsei
University, Seoul 03722, Korea}
\author{Sungsik Noh}
\email{sungsiknoh@yonsei.ac.kr}\affiliation{Department of Physics and Institute of Physics and Applied Physics, Yonsei
University, Seoul 03722, Korea}
\author{Su Houng Lee}
\email{suhoung@yonsei.ac.kr}\affiliation{Department of Physics and Institute of Physics and Applied Physics, Yonsei
University, Seoul 03722, Korea}

\begin{abstract}
We perform a constituent quark model analysis for the masses of the doubly heavy tetraquark states $T_{QQ}$ after we fix the parameters to fit the masses of the newly observed $\Xi_{cc}^{++}$ and  hadrons involving heavy quarks relevant to the stability of these states.  
We investigate in detail how the relative distances between quark pairs vary as we change the quark content and how they affect the various contributions to the total tetraquark masses. 
We also find that our full calculations give in general less binding compared to simplified quark model calculations that 
treat quark dynamics inside the tetraquark the same as that inside a baryon.   
We trace the main origin to be the differences in the number of relative kinetic energies which increases as one goes from meson, baryon and tetraquarks.
   We also compare our new results with previous works using less constrained parameters and find that the tetraquark state $T_{bb}(ud\bar{b}\bar{b})$ and $T_{bb}(us\bar{b}\bar{b})$ are bound by 120.56 MeV and 7.3 MeV respectively. 

\end{abstract}

\maketitle

\section{ INTRODUCTION}

With the recent discovery of the doubly charmed baryon $\Xi_{cc}^{++}$\cite{Aaij:2017ueg,Aaij:2018wzf} and its decay\cite{Aaij:2018gfl}, there is currently a great interest in doubly charmed hadrons\cite{Chen:2017sbg,Karliner:2017qjm,Francis:2016hui,Hong:2018mpk}.  While there were previous reports on the observation of a doubly charmed baryon\cite{Ocherashvili:2004hi}, the present observation marks the first unambiguous observation of the doubly heavy baryon with the measured  mass different from the earlier claim.  
The existence of the doubly charmed baryon was expected, but the observation of the $\Xi_{cc}^{++}$ makes it possible to quantitatively analyze hadron configurations with two heavy quarks.  Such information was recently used in a simplified quark based model to asses the stability of doubly heavy tetraquark states\cite{Karliner:2017qjm}, where  it was speculated for a long time that a stable flavor exotic meson could exist\cite{Zouzou:1986qh,Lipkin:1986dw,Manohar:1992nd,Brink:1998}.  

Over the years, the observation of several exotic candidates with heavy quarks were reported.  These started from the observation of $D_{sJ}(2317)$
\cite{Aubert:2003fg}, $X(3872)$ \cite{Choi:2003ue} and
continues to the present day with the recent observation of
$P_c(4380)^+$ and $P_c(4450)^+$ \cite{Aaij:2015tga}.  
However, there is a special interest in the doubly heavy tetraquark
$T_{QQ}(QQ\bar{u}\bar{d})$ state with isospin zero and in particular with the quantum number
$I(J^P)=0(1^+)$ \cite{Zouzou:1986qh,Lipkin:1986dw,Manohar:1992nd}.
First of all, this particle is  a flavor exotic tetraquark, which has never
been observed before. Second, with the discovery of the
doubly charmed baryon\cite{Aaij:2017ueg}, the chances 
of observing a similar doubly charmed hadron with the light quark
replaced by a strongly correlated light anti-diquark seems quite possible. Finally, analyzing the structure of this particle in
the constituent quark model, one finds that this particle is the
only candidate where there is a strong attraction in the compact
configuration compared to two separated mesons. This is so because
while previously observed exotic candidates such as the $X(3872)$
is composed of $q\bar{q} Q\bar{Q}$, where $q,Q$ are light and
heavy quarks respectively, the proposed $T_{QQ}$ state is composed
of $QQ \bar{q} \bar{q}$ quarks. The latter quark structure favours
a compact tetraquark configuration as the additional light anti-diquark structure
$\bar{q} \bar{q}$ in the isospin zero channel provides an
attraction larger than that for  the two $Q \bar{q}$ in a
separated meson configuration
\cite{Park:2013fda,Ohkoda:2012hv,Hyodo:2017hue,Luo:2017eub}. Hence, the $T_{QQ}$ is
a unique multiquark candidate state that could be compact.  In fact, recent lattice calculations also show a short range attraction in the $T_{QQ}$ channel\cite{Ikeda:2013vwa,Bicudo:2015vta}.

In this work, we will perform a detailed constituent quark model analysis of the tetraquark state after we fix the parameters to fit the masses of hadrons involving heavy quarks relevant to the stability of $T_{QQ}$.  Up to now, no hadrons with two heavy quarks were found so that fitting the mass of $\Xi_{cc}^{++}$  provides crucial input to study other configurations involving two heavy quarks.  Although the authors of Ref.~\cite{Karliner:2014gca}  predicted the mass of $\Xi_{cc}$ with good accuarcy using a simplified quark model and recently reanalyzed the stability of $T_{QQ}$ within their model\cite{Karliner:2017qjm}, it should be noted that a detailed analysis is necessary and important in the quark model when analysing compact multiquark configurations as one has to systematically understand not only the full color spin flavor wave function but also the full quark interactions in a multiquark configuration. 
The reason why quarks have to be treated carefully  in a multiquark configuration  is already evident from noting the need to introduce  
different constituent quark masses in the meson and baryon   to fit the hadron spectrum  in a simplified quark model\cite{Karliner:2014gca}.  As we will see later, the charm (light) quark interactions  inside the $\Xi_{cc}$ ($\Lambda_c$) cannot be universally translated to those inside $T_{QQ}$.
The detailed analysis of the quark dynamics in a multquark configuration is essential to understand the stability of a compact configuration against the lowest threshold hadronic states.   Such an analysis for a compact dibaryon configuration indicated that the recently discovered $d^*(2380)$\cite{Bashkanov:2008ih} cannot be a compact multiquark configuration, which suggests that it would be a molecular configurations composed of well separated baryons\cite{Park:2015nha}.

In section II, we first introduce the constituent quark model and the fitting of the  parameters.  
 In section III, the spatial wave function and the flavor color spin wave functions are introduced.  
In section IV, the numerical results are presented. In particular, we will show in detail the relative distances between quark pairs in all configurations and how they are related to individual contributions to the total mass. Furthermore, we will also present a detailed comparison of our model result to that from a simplified quark model. Discussion and summary are given in section V.

\section{Formalism}

We use a nonrelativistic Hamiltonian for the consitutent quarks of the following form \cite{Brink:1998}.
\begin{eqnarray}
H &=& \sum^{4}_{i=4} \left( m_i+\frac{{\mathbf p}^{2}_i}{2 m_i} \right)-\frac{3}{4}\sum^{4}_{i<j}\frac{\lambda^{c}_{i}}{2} \cdot \frac{\lambda^{c}_{j}}{2} \left( V^{C}_{ij} + V^{CS}_{ij} \right), \qquad
\label{hamiltonian}
\end{eqnarray}
where $m_i$ are the quark masses and $\lambda^c_{i}/2$ is the color operator of the $i$-th quark for the color SU(3). For the internal quark potentials $V^{C}_{ij}$ and $V^{CS}_{ij}$, we adopt the following forms \cite{Brink:1998,Woosung:2017}: 
\begin{eqnarray}
V^{C}_{ij} &=& - \frac{\kappa}{r_{ij}} + \frac{r_{ij}}{a^2_0} - D,
\label{potential1}
\\
V^{CS}_{ij} &=& \frac{\hbar^2 c^2 \kappa'}{m_i m_j c^4} \frac{e^{- \left( r_{ij} \right)^2 / \left( r_{0ij} \right)^2}}{(r_{0ij}) r_{ij}} \sigma_i \cdot \sigma_j.
\label{potential2}
\end{eqnarray}
Here
\begin{eqnarray}
r_{0ij} &=& 1/ \left( \alpha + \beta \frac{m_i m_j}{m_i + m_j} \right)	,	
\label{potential3}
\\
\kappa' &=& \kappa_0 \left( 1+ \gamma \frac{m_i m_j}{m_i + m_j} \right)	,	
\label{potential4}
\end{eqnarray}
where $r_{ij}=|{\mathbf r}_i - {\mathbf r}_j |$ is the distance between the quarks $i$ and $j$ of masses $m_i$ and $m_j$, respectively, and $\sigma_i$ is the spin operator. For the parameters appearing in Eqs.~(\ref{potential1})-(\ref{potential4}), we fit them to the experimental meson and baryon masses listed in Tables~\ref{meson}, \ref{baryon}. 
The wave functions for the baryons and mesons are chosen as below.
\begin{eqnarray}
R_{Meson} & = & \exp [ -C \sigma^2],  \nonumber \\
R_{Baryon} & = & \exp  [ - C_{11}^s \sigma^2 -C_{22}^s \lambda^2],
\end{eqnarray}
where ${\boldsymbol \sigma} = \frac{1}{\sqrt{2}} \left( {\mathbf r}_1 - {\mathbf r}_2 \right)$ for mesons, and
\begin{small}
\begin{eqnarray}
&&{\boldsymbol \sigma} = \frac{1}{\sqrt{2}} \left( {\mathbf r}_1 - {\mathbf r}_2 \right)	\nonumber
\\
&&{\boldsymbol \lambda} = [2 (m_1^2 + m_2^2 + m_1 m_2)]^{-1/2} \left[ m_1 {\mathbf r}_1 + m_2 {\mathbf r}_2 - (m_1 + m_2) {\mathbf r}_3 \right]	\nonumber
\end{eqnarray}
\end{small}
for baryons in our reference of calculations. Using variational method, the parameters are then determined as follows:
\begin{eqnarray}
&\kappa=120.0 \, \textrm{MeV fm}, \quad a_0=0.0318119 \, \textrm{(MeV$^{-1}$fm)$^{1/2}$}, & \nonumber \\
& D=983  \, \textrm{MeV}, & \nonumber \\
&m_{u}=326 \, \textrm{MeV}, \qquad m_{s}=607 \, \textrm{MeV}, &\nonumber \\
&m_{c}=1918 \, \textrm{MeV}, \qquad m_{b}=5323 \, \textrm{MeV},	&\nonumber \\
&\alpha = 1.0499 \, \textrm{fm$^{-1}$}, \,\, \beta = 0.0008314 \, \textrm{(MeV fm)$^{-1}$}, &	\nonumber \\
&\gamma = 0.00088 \, \textrm{MeV$^{-1}$}, \,\, \kappa_0=194.144 \, \textrm{MeV}. 	 &
\label{parameters}
\end{eqnarray}

\begin{table}[t]

\caption{The masses of mesons containing at least one $c$ or $b$ quark. The unit of the variational parameter $C$ is fm$^{-2}$.}	
\centering
\begin{tabular}{ccccc}
\hline
\hline \, \multirow{2}{*}{Particle} \, & \,\,\, Mass \,\,\, & \, Variational \, & \, Experimental \, & \,\, Error \\
       \,          \, & \,\,\, (MeV) \,\,\, & \,  Parameter $C$  \, & \,     Value (MeV)   \, & \,\,  (\%) \\
\hline 
\\
$\eta_{c}$		&	2998.5		&	15.0	&	2983.6		& 0.50		\\
$J/\psi$		&	3092.22		&	12.5	&	3096.92		& 0.15		\\
$\eta_{b}$		&	9351.2		&	57.1	&	9398.0		& 0.50		\\
$\Upsilon$		&	9430.97		&	49.4	&	9460.30		& 0.31		\\
$\bar{D}^{0}$	&	1869.35		&	4.6		&	1864.84		& 0.24		\\
$\bar{D}^{*0}$	&	1996.93		&	3.8		&	2006.96		& 0.50		\\
$B^{+}$			&	5294.50		&	4.6		&	5279.26		& 0.29		\\
$B^{*+}$		&	5343.7		&	4.2		&	5325.2		& 0.35		\\
$B^{0}_{s}$	&	5347.99		&	7.8		&	5366.77		& 0.35		\\
$B^{*0}_{s}$	&	5397.1		&	7.1		&	5415.4		& 0.34		\\
\\
\hline 
\hline
\label{meson}
\end{tabular}
\end{table}

\begin{table}[t]
\caption{The masses of baryons  containing at least one $c$ or $b$ quark.}	
\centering
\begin{tabular}{ccccc}
\hline
\hline \multirow{2}{*}{Particle} & \, Mass \qquad & \quad Variational \quad & Experimental \,\, & Error \\
                & \, (MeV) \qquad & \quad Parameters (fm$^{-2}$) \quad  &    Value (MeV) \,\,  & (\%) \\
\hline 
\\
$\Lambda^{0}$		& 1110.43	& \, $C^{s}_{11}=2.7, C^{s}_{22}=2.7$	& 1115.68 & 0.31	\\
$\Lambda^{0}_{b}$	& 5637.1	& \, $C^{s}_{11}=2.9, C^{s}_{22}=4.2$	& 5619.4 & 0.31	\\
$\Lambda^{+}_{c}$	& 2283.21	& \, $C^{s}_{11}=2.8, C^{s}_{22}=3.7$	& 2286.46 & 0.14 	\\
$\Sigma^{++}_{c}$	& 2445.18	& \, $C^{s}_{11}=2.1, C^{s}_{22}=3.7$	& 2453.98 & 0.36	\\
$\Sigma^{*++}_{c}$& 2518.3	& \, $C^{s}_{11}=2.0, C^{s}_{22}=3.4$	& 2517.9 & 0.02 	\\
$\Sigma^{+}_{b}$	& 5832.1	& \, $C^{s}_{11}=2.1, C^{s}_{22}=4.1$	& 5811.3 &	0.36	\\
$\Sigma^{*+}_{b}$	& 5861.0	& \, $C^{s}_{11}=2.0, C^{s}_{22}=3.9$	& 5832.1 & 0.50	\\
$\Xi^{++}_{cc}$	& 3612.37	& \, $C^{s}_{11}=8.0, C^{s}_{22}=3.2$	& 3621.40 & 0.25	\\
\\
\hline 
\hline
\label{baryon}
\end{tabular}
\end{table}

We emphasize that it is important to fit the paramters also to the recently measured $\Xi_{cc}$ mass.  Using the parameters used  before\cite{Brink:1998,Park:2013fda}, one finds the mass of $\Xi_{cc}$ to be 3653.30 MeV, which is substantially larger than the experimentally measured value.  In fact, the previous parameter set tends to systematically give slightly larger heavy meson mass and smaller hyperfine mass splitting.  Furthermore, it is important to introduce additional mass dependence in the coefficients for the hyperfine mass splitting such as in Eq.~(\ref{potential3})-(\ref{potential4}) so that the hyperfine splitting between heavy quarks are larger than the naive extrapolation from the corresponding splitting between lighter quarks obtained with the constituent quark masses appearing only as an overall factor in the denominator as in Eq.~(\ref{potential2}).

\section{WAVE FUNCTION}

In the constituent quark model, the lowest mass is obtained in a configuration where all the quarks are in the $l=0$ state. Thus, the Hamiltonian introduced in Eq.~(\ref{hamiltonian}) will be applied only to the $s$-wave configurations. Here, we present the wave function of the spatial, color-spin, and flavor structure of a tetraquark system.

\subsection{Spatial Function}

We construct the trial wave function for the spatial part in a simple Gaussian form as in the previous work\cite{Park:2013fda}. In the case where the constituent quark masses are all different, the Jacobi coordinates are as follows:
\begin{itemize}
  \item{Coordinate 1}:
  \begin{eqnarray}
  & \boldsymbol{\rho} = \frac{1}{\sqrt{2}}({\mathbf r}_1 - {\mathbf r}_3), \qquad \boldsymbol{\rho}' = \frac{1}{\sqrt{2}}({\mathbf r}_4 - {\mathbf r}_2) &	\nonumber
  \\
  & {\mathbf x} = \frac{1}{\mu} \left( \frac{m_1 {\mathbf r}_1 + m_2 {\mathbf r}_3}{m_1 + m_2} - \frac{m_3 {\mathbf r}_2 + m_4 {\mathbf r}_4}{m_3 + m_4} \right) &
  \end{eqnarray}
  \item{Coordinate 2}:
  \begin{eqnarray}
  & \boldsymbol{\alpha} = \frac{1}{\sqrt{2}}({\mathbf r}_1 - {\mathbf r}_4), \qquad \boldsymbol{\alpha}' = \frac{1}{\sqrt{2}}({\mathbf r}_2 - {\mathbf r}_3) &	\nonumber
  \\
  & {\mathbf y} = \frac{1}{\mu} \left( \frac{m_1 {\mathbf r}_1 + m_2 {\mathbf r}_4}{m_1 + m_2} - \frac{m_3 {\mathbf r}_2 + m_4 {\mathbf r}_3}{m_3 + m_4} \right) &
  \end{eqnarray}
  \item{Coordinate 3}:
  \begin{eqnarray}
  & \boldsymbol{\sigma} = \frac{1}{\sqrt{2}}({\mathbf r}_1 - {\mathbf r}_2), \qquad \boldsymbol{\sigma}' = \frac{1}{\sqrt{2}}({\mathbf r}_3 - {\mathbf r}_4) &	\nonumber
  \\
  & \boldsymbol{\lambda} = \frac{1}{\mu} \left( \frac{m_1 {\mathbf r}_1 + m_2 {\mathbf r}_2}{m_1 + m_2} - \frac{m_3 {\mathbf r}_3 + m_4 {\mathbf r}_4}{m_3 + m_4} \right) &
  \end{eqnarray}
\end{itemize}
where
\begin{eqnarray}
\mu &=& \left[ \frac{m_1^2 + m_2^2}{(m_1 + m_2)^2} + \frac{m_3^2 + m_4^2}{(m_3 + m_4)^2} \right]^{1/2}	\nonumber
\end{eqnarray}
We set the Jacobi coordinates with the following conditions:
\begin{eqnarray}
&m_u=m_d;&	\nonumber
\\
&m_1=m_2=m_u,	\,\,	m_3=m_4=m_c& 	\quad {\rm for} \,\, ud\bar{c}\bar{c} ,	\nonumber
\\
&m_1=m_2=m_u,	\,\,	m_3=m_4=m_b& 	\quad {\rm for} \,\, ud\bar{b}\bar{b},	\nonumber
\\
&m_1=m_2=m_u,	\,\,	m_3=m_c, \,\, m_4=m_b& 	\quad {\rm for} \,\, ud\bar{c}\bar{b},	\nonumber
\\
&m_1=m_u, \,\, m_2=m_s,	\,\,	m_3=m_4=m_b& 	\quad {\rm for} \,\, us\bar{b}\bar{b}.	\nonumber
\end{eqnarray}

As discussed in Ref.~\cite{Brink:1998}, the most general Gaussian form for the $s$-wave spatial function $R^s$ can be written as follows.
\begin{widetext}
\begin{small}
\begin{eqnarray}
&R^s = {\rm exp} \big[-(A^s_{11} \rho^2 + A^s_{22} \rho'^2 + A^s_{33} x^2 + 2 A^s_{12} \boldsymbol{\rho} \cdot \boldsymbol{\rho}' + 2 A^s_{13} \boldsymbol{\rho} \cdot {\mathbf x} + 2 A^s_{23} \boldsymbol{\rho}' \cdot {\mathbf x} ) \big]
={\rm exp} \big[ - (B^s_{11} \alpha^2 + B^s_{22} \alpha'^2 + B^s_{33} y^2 + 2B^s_{12} \boldsymbol{\alpha} \cdot \boldsymbol{\alpha}' &	\nonumber
\\
&+ 2 B^s_{13} \boldsymbol{\alpha} \cdot {\mathbf y} + 2 B^s_{23} \boldsymbol{\alpha}' \cdot {\mathbf y} ) \big]
={\rm exp} \big[ - (C^s_{11} \sigma^2 + C^s_{22} \sigma'^2 + C^s_{33} \lambda^2 + 2C^s_{12} \boldsymbol{\sigma} \cdot \boldsymbol{\sigma}' + 2 C^s_{13} \boldsymbol{\sigma} \cdot \boldsymbol{\lambda} + 2 C^s_{23} \boldsymbol{\sigma}' \cdot \boldsymbol{\lambda} ) \big]. & \label{spatial-wave}
\end{eqnarray}
\end{small}
\end{widetext}
For symmetry reason, we take the coordinate set 3 as our reference. Also, the calculation of the spatial function will be performed for the case where $C^s_{12}=C^s_{13}=C^s_{23}=0$ so that the non-zero  variational parameters are $C^s_{11}$, $C^s_{22}$, $C^s_{33}$.
\\

At the same time, it is useful to introduce the center of mass frame so that the kinetic term in the Hamiltonian in Eq.~(\ref{hamiltonian}) can be reduced appropriately for our calculations. The kinetic term in the center of mass frame denoted by $T_c$ is as follows.
\begin{eqnarray}
T_c &=& \sum^4_{i=1} \frac{{\mathbf p}^2_i}{2m_i} - \frac{{\mathbf p}^2_{rC}}{2M}
=\frac{{\mathbf p}^2_{\mathbf \sigma}}{2m'_1} + \frac{{\mathbf p}^2_{\mathbf \sigma'}}{2m'_2} + \frac{{\mathbf p}^2_{\mathbf \lambda}}{2m'_3}
\end{eqnarray}
where
\begin{eqnarray}
&&m'_1=m_u, \,\, m'_2=m_c, \,\, m'_3=\frac{2 m_u m_c}{m_u + m_c}	\quad {\rm for} \,\, ud\bar{c}\bar{c},	\nonumber
\end{eqnarray}
\begin{eqnarray}
&&m'_1=m_u, \,\, m'_2=m_b, \,\, m'_3=\frac{2 m_u m_b}{m_u + m_b}	\quad {\rm for} \,\, ud\bar{b}\bar{b},	\nonumber
\\
&&m'_1=m_u, \,\, m'_2=\frac{2 m_c m_b}{m_c + m_b}, 	\nonumber \\
&&\hspace{0.28cm} m'_3=\frac{(3 m_c^2 + 2 m_c m_b + 3 m_b^2) m_u}{(m_c + m_b) (2 m_u + m_c + m_b)}  \,\,\quad {\rm for} \,\, ud\bar{c}\bar{b},	\nonumber	\\
&&m'_1=\frac{2 m_u m_s}{m_u + m_s}, \,\, m'_2=m_b, 	\nonumber \\
&&\hspace{0.28cm} m'_3=\frac{(3 m_u^2 + 2 m_u m_s + 3 m_s^2) m_b}{(m_u + m_s) (m_u + m_s + 2 m_b)}  \,\,\quad {\rm for} \,\, us\bar{b}\bar{b}.	\nonumber
\\
\label{masses}
\end{eqnarray}

\subsection{Flavor Part}

In flavor SU(2), the subparticle($ud$) system can be in one of the two representation as denoted by the Young tableaux on the R.H.S.
\\

\hspace{1.88cm}
\begin{small}
$\begin{tabular}{|c|}
  \hline
  \quad \quad  \\
  \hline
\end{tabular}$ $\otimes$
$\begin{tabular}{|c|}
  \hline
  \quad \quad  \\
  \hline
\end{tabular}$
=
$\begin{tabular}{|c|}
  \hline
  \quad \quad  \\
  \hline
  \quad \quad  \\
  \hline
\end{tabular}$ $\oplus$
$\begin{tabular}{|c|c|}
  \cline{1-2}
  \quad \quad & \quad \quad \\
  \cline{1-2}
\end{tabular}$
\end{small}

{}\
\\
{}\
\\
Thus, in order to satisfy the Pauli's principle,  the color-spin part of the wave function for the $I=0$ configuration should be symmetric under the transposition between the quarks $u$ and $d$, while for the  $I=1$ state, it should be antisymmetric. Since the quarks $c$ and $b$ have isospin 0, for the subparticle ($\bar{c}\bar{c}$ or $\bar{b}\bar{b}$), the color-spin part should be antisymmetric under the corresponding transposition. For the other combinations of subparticle systems such as ($us$) or ($\bar{c}\bar{b}$), they are different particles so that there is no such a symmetry constraint on the construction of the total wave function.

\subsection{Color-Spin Part}

The color space in a given flavor configuration of the tetraquarks ($qq\bar{Q}\bar{Q}$) can be decomposed in terms of the color SU(3).
\begin{eqnarray}
\hspace{-0.9cm}&&[3]_c \otimes [3]_c \otimes [\bar{3}]_c \otimes [\bar{3}]_c 	\nonumber
\\
\hspace{-0.9cm}&&=
[\bar{3}]_c \otimes [3]_c \oplus [6]_c \otimes [\bar{6}]_c \oplus [\bar{3}]_c \otimes [\bar{6}]_c \oplus [6]_c \otimes [3]_c
\end{eqnarray}
Color singlet states can be obtained from the first and the second terms on the R.H.S. It is convenient to use the following notation\cite{Buccella:2007} to denote the two color siglets:
\begin{eqnarray}
\left( q_1 q_2 \right)^{\bar{3}} \otimes \left( \bar{q}_3 \bar{q}_4 \right)^{3}, \qquad \left( q_1 q_2 \right)^{6} \otimes \left( \bar{q}_3 \bar{q}_4 \right)^{\bar{6}}.
\end{eqnarray}
The symmetry property of the two color siglets under the transpositions are as follows.
\begin{eqnarray}
\hspace{-0.9cm}(12) \left( q_1 q_2 \right)^{\bar{3}} \otimes \left( \bar{q}_3 \bar{q}_4 \right)^{3}
&=&
(34) \left( q_1 q_2 \right)^{\bar{3}} \otimes \left( \bar{q}_3 \bar{q}_4 \right)^{3}
\nonumber
\\
&=& -\left( q_1 q_2 \right)^{\bar{3}} \otimes \left( \bar{q}_3 \bar{q}_4 \right)^{3},
\\
\hspace{-0.9cm}(12) \left( q_1 q_2 \right)^{6} \otimes \left( \bar{q}_3 \bar{q}_4 \right)^{\bar{6}}
&=&
(34) \left( q_1 q_2 \right)^{6} \otimes \left( \bar{q}_3 \bar{q}_4 \right)^{\bar{6}}
\nonumber
\\
&& \left( q_1 q_2 \right)^{6} \otimes \left( \bar{q}_3 \bar{q}_4 \right)^{\bar{6}}.
\end{eqnarray}
The two color singlets can be written as:
\begin{eqnarray}
\hspace{-1.3cm}&&\left( q_1 q_2 \right)^{\bar{3}} \otimes \left( \bar{q}_3 \bar{q}_4 \right)^{3}	\nonumber
\\
\hspace{-1.3cm}&&\hspace{0.7cm}=
\frac{1}{\sqrt{12}}\epsilon^{\alpha \beta \gamma} \epsilon_{\alpha \lambda \sigma} q_{\beta}(1)q_{\gamma}(2)\bar{q}^{\lambda}(3)\bar{q}^{\sigma}(4),	\nonumber
\\
\hspace{-1.3cm}&&\left( q_1 q_2 \right)^{6} \otimes \left( \bar{q}_3 \bar{q}_4 \right)^{\bar{6}}
\nonumber
\\
\hspace{-1.3cm}&&\hspace{0.7cm}=
\frac{1}{\sqrt{6}}d^{\alpha \beta \gamma} d_{\alpha \lambda \sigma} q_{\beta}(1)q_{\gamma}(2)\bar{q}^{\lambda}(3)\bar{q}^{\sigma}(4),
\end{eqnarray}
where $d^{\alpha \beta \gamma}$ and $d_{\alpha \beta\ \gamma}$ are
\begin{eqnarray}
d^{111}&=&d_{111}=d^{222}=d_{222}=d^{333}=d_{333}=1,	\nonumber
\\
d^{412}&=&d_{412}=d^{421}=d_{421}=d^{523}=d_{523}=d^{532}=d_{532}	\nonumber
\\
&=&d^{613}=d_{613}=d^{631}=d_{631}=\frac{1}{\sqrt{2}}.
\end{eqnarray}

From the definitions of $\epsilon^{\alpha \beta \gamma}$ and $d^{\alpha \beta \gamma}$, we recognize that the color singlet state of the tetraquark has the internal structure consisting of two quark-antiquark pairs. On the other hand, the two color singlet states can be recombined into another two color singlets constructed from two quark-antiquark pairs of color singlet-singlet and an octet-octet states which are appropriate for studying the decay properties.
\begin{eqnarray}
&&\left( q_1 \bar{q}_3 \right)^{1} \otimes \left( q_2 \bar{q}_4 \right)^{1}	
=
\frac{1}{3} q_{\alpha}(1)\bar{q}^{\alpha}(3) q_{\beta}(2)\bar{q}^{\beta}(4),	\nonumber
\\
&&\left( q_1 \bar{q}_3 \right)^{8} \otimes \left( q_2 \bar{q}_4 \right)^{8}	\nonumber
\\
&&=
\frac{1}{2\sqrt{2}} \left( q_{\alpha}(1)\bar{q}^{\alpha}(4) q_{\beta}(2)\bar{q}^{\beta}(3) - \frac{1}{3} q_{\alpha}(1)\bar{q}^{\alpha}(3) q_{\beta}(2)\bar{q}^{\beta}(4) \right),	\nonumber
\end{eqnarray}
or
\begin{eqnarray}
&&\left( q_1 \bar{q}_4 \right)^{1} \otimes \left( q_2 \bar{q}_3 \right)^{1}	
=
\frac{1}{3} q_{\alpha}(1)\bar{q}^{\alpha}(4) q_{\beta}(2)\bar{q}^{\beta}(3),	\nonumber
\\
&&\left( q_1 \bar{q}_4 \right)^{8} \otimes \left( q_2 \bar{q}_3 \right)^{8}	\nonumber
\\
&&=
\frac{1}{2\sqrt{2}} \left( q_{\alpha}(1)\bar{q}^{\alpha}(3) q_{\beta}(2)\bar{q}^{\beta}(4) - \frac{1}{3} q_{\alpha}(1)\bar{q}^{\alpha}(4) q_{\beta}(2)\bar{q}^{\beta}(3) \right).	\nonumber
\end{eqnarray}

In the fundamental representation of SU(2), the spin space of the tetraquark can be represented as $[2] \otimes [2] \otimes [2] \otimes [2]$ and decomposed into the direct sums by means of the bases of the permutation group. The decomposition can be represented as Young tableaux.
\\

\hspace{0.1cm}
\begin{small}
$\begin{tabular}{|c|}
  \hline
  \quad \quad  \\
  \hline
\end{tabular}$ $\otimes$
$\begin{tabular}{|c|}
  \hline
  \quad \quad  \\
  \hline
\end{tabular}$ $\otimes$
$\begin{tabular}{|c|}
  \hline
  \quad \quad  \\
  \hline
\end{tabular}$ $\otimes$
$\begin{tabular}{|c|}
  \hline
  \quad \quad  \\
  \hline
\end{tabular}$
\\

\hspace{0.5cm}=
$\begin{tabular}{|c|c|c|c|}
  \cline{1-4}
  \quad \quad & \quad \quad & \quad \quad & \quad \quad \\
  \cline{1-4}
\end{tabular}$ $\oplus$
$\begin{tabular}{|c|c|}
  \cline{1-2}
  \quad \quad & \quad \quad \\
  \cline{1-2}
  \quad \quad & \quad \quad \\
  \cline{1-2}
\end{tabular}$ $\oplus$
$\begin{tabular}{|c|c|}
  \cline{1-2}
  \quad \quad & \quad \quad \\
  \cline{1-2}
  \quad \quad & \quad \quad \\
  \cline{1-2}
\end{tabular}$
\\
\\

\hspace{0.8cm}
$\oplus$
$\begin{tabular}{|c|c|c|}
  \cline{1-3}
  \quad \quad & \quad \quad & \quad \quad \\
  \cline{1-3}
  \quad \quad	\\
  \cline{1-1}
\end{tabular}$ $\oplus$
$\begin{tabular}{|c|c|c|}
  \cline{1-3}
  \quad \quad & \quad \quad & \quad \quad \\
  \cline{1-3}
  \quad \quad	\\
  \cline{1-1}
\end{tabular}$ $\oplus$
$\begin{tabular}{|c|c|c|}
  \cline{1-3}
  \quad \quad & \quad \quad & \quad \quad \\
  \cline{1-3}
  \quad \quad	\\
  \cline{1-1}
\end{tabular}$,
\end{small}
{}\
\\
\\
where the number of each shape of diagram indicates the number of independent bases for the corresponding spin space. With the dimensions of the spin SU(2) for each diagram, the decomposition can be written as
\begin{eqnarray}
[2] \otimes [2] \otimes [2] \otimes [2]
=
[5] \oplus [3] \oplus [3] \oplus [3] \oplus [1] \oplus [1]	\nonumber
\end{eqnarray}
Thus, the number of independent bases for each spin space is 1, 3, and 2 for spin 2, 1, 0 spaces, respectively.
\\
\\

\noindent (a) {\it Spin 0 case}
\\
There are two independent basis states obtained from the combinations of two spin 0 subparticles and of two spin 1 subparticles, which are denoted\cite{Park:2013fda} by
\begin{eqnarray}
\left( \chi_{12} \right)_{s=0} \otimes \left( \chi_{34} \right)_{s=0}, \quad \left( \chi_{12} \right)_{s=1} \otimes \left( \chi_{34} \right)_{s=1}.
\end{eqnarray}
The symmetry property under the transpositions are given by
\begin{eqnarray}
\hspace{-0.8cm}(12)\left( \chi_{12} \right)_{s=0} \otimes \left( \chi_{34} \right)_{s=0} &=& (34)\left( \chi_{12} \right)_{s=0} \otimes \left( \chi_{34} \right)_{s=0}	\nonumber
\\
&=&-\left( \chi_{12} \right)_{s=0} \otimes \left( \chi_{34} \right)_{s=0},	\nonumber
\\
\hspace{-0.8cm}(12)\left( \chi_{12} \right)_{s=1} \otimes \left( \chi_{34} \right)_{s=1} &=& (34)\left( \chi_{12} \right)_{s=1} \otimes \left( \chi_{34} \right)_{s=1}	\nonumber
\\
&=&\left( \chi_{12} \right)_{s=1} \otimes \left( \chi_{34} \right)_{s=1}.
\end{eqnarray}

\begin{widetext}

\begin{table}[t]
\caption{The mass and binding energy $B_{T}$ for each tetraquark state.}	
\begin{tabular}{cccccc}
\hline
\hline \quad Type \quad  & \quad ($I$, $S$) \quad & \quad Color-Spin Bases \quad & \quad Mass(MeV) \quad & \quad \, Variational Parameters(fm$^{-2}$) \quad & \quad \, $B_{T}$(MeV) \\
\hline 
	\\
$ud\bar{b}\bar{b}$ & \, (0,1) &  $\psi_3, \,\, \psi_6$  & 10517.67 & \quad $C^{s}_{11}=2.8, \, C^{s}_{22}=20.9, \, C^{s}_{33}=2.8$ & \quad -120.56 \\
 & \, (1,0) & $\phi_2, \,\, \phi_3$ & 10710.15 & \quad $C^{s}_{11}=2.1, \, C^{s}_{22}=20.7, \, C^{s}_{33}=2.8$ & \quad +121.15 \\
 & \, (1,1) & $\psi_2$ & 10722.23 & \quad $C^{s}_{11}=2.1, \, C^{s}_{22}=20.8, \, C^{s}_{33}=2.7$ & \quad +84.0 \\
 \\
$ud\bar{c}\bar{c}$ & \, (0,1) & $\psi_3, \,\, \psi_6$ & 3964.90 & \quad $C^{s}_{11}=2.8, \, C^{s}_{22}=7.6, \, C^{s}_{33}=2.7$ & \quad +98.62 \\
 & \, (1,0) & $\phi_2, \,\, \phi_3$ & 4103.82 & \quad $C^{s}_{11}=2.1, \, C^{s}_{22}=7.0, \, C^{s}_{33}=3.1$ & \quad +365.12 \\
 & \, (1,1) & $\psi_2$ & 4158.31 & \quad $C^{s}_{11}=2.0, \, C^{s}_{22}=7.6, \, C^{s}_{33}=2.5$ & \quad +292.03 \\
\\
$ud\bar{c}\bar{b}$ & \, (0,0) & $\phi_1, \,\, \phi_4$ & 7238.08 & \quad $C^{s}_{11}=3.1, \, C^{s}_{22}=10.8, \, C^{s}_{33}=2.8$ & \quad +74.23 \\
 & \, (0,1) & $\psi_1, \,\, \psi_3, \,\, \psi_6$ & 7262.45 & \quad $C^{s}_{11}=3.1, \, C^{s}_{22}=10.3, \, C^{s}_{33}=2.7$ & \quad +49.37 \\
 & \, (1,0) & $\phi_2, \,\, \phi_3$ & 7454.79 & \quad $C^{s}_{11}=2.3, \, C^{s}_{22}=10.1, \, C^{s}_{33}=2.7$ & \quad +290.94 \\
 & \, (1,1) & $\psi_2, \,\, \psi_4, \,\, \psi_5$ & 7479.05 & \quad $C^{s}_{11}=2.3, \, C^{s}_{22}=10.2, \, C^{s}_{33}=2.6$ & \quad +265.97 \\
 \\
$us\bar{b}\bar{b}$ & \, (1/2,0) & $\phi_2, \,\, \phi_3$ & 10814.4 & \quad $C^{s}_{11}=2.9, \, C^{s}_{22}=20.4, \, C^{s}_{33}=3.5$ & \quad +171.91 \\
& \, (1/2,1) & $\psi_2, \,\, \psi_3, \,\, \psi_6$ & 10684.4 & \quad $C^{s}_{11}=3.5, \, C^{s}_{22}=20.6, \, C^{s}_{33}=3.5$ & \quad -7.3 \\
\\
\hline 
\hline
\label{result}
\end{tabular}
\end{table}

\end{widetext}

\noindent (b) {\it Spin 1 case}
\\
The three independent bases are given by
\begin{eqnarray}
&\left( \chi_{12} \right)_{s=0} \otimes \left( \chi_{34} \right)_{s=1}, \quad \left( \chi_{12} \right)_{s=1} \otimes \left( \chi_{34} \right)_{s=0},&	\nonumber
\\
&\left( \chi_{12} \right)_{s=1} \otimes \left( \chi_{34} \right)_{s=1}. &
\end{eqnarray}
The symmetry property under the transpositions are given by
\begin{eqnarray}
\hspace{-0.6cm}(12)\left( \chi_{12} \right)_{s=1} \otimes \left( \chi_{34} \right)_{s=0} &=& -(34)\left( \chi_{12} \right)_{s=1} \otimes \left( \chi_{34} \right)_{s=0}	\nonumber
\\
&=&\left( \chi_{12} \right)_{s=1} \otimes \left( \chi_{34} \right)_{s=0},	\nonumber
\\
\hspace{-0.6cm}(12)\left( \chi_{12} \right)_{s=1} \otimes \left( \chi_{34} \right)_{s=1} &=& (34)\left( \chi_{12} \right)_{s=1} \otimes \left( \chi_{34} \right)_{s=1}	\nonumber
\\
&=&\left( \chi_{12} \right)_{s=1} \otimes \left( \chi_{34} \right)_{s=1},	\nonumber
\\
\hspace{-0.6cm}(12)\left( \chi_{12} \right)_{s=0} \otimes \left( \chi_{34} \right)_{s=1} &=& -(34)\left( \chi_{12} \right)_{s=0} \otimes \left( \chi_{34} \right)_{s=1}	\nonumber
\\
&=&-\left( \chi_{12} \right)_{s=0} \otimes \left( \chi_{34} \right)_{s=1}.
\end{eqnarray}

\noindent (c) {\it Spin 2 case}
\\
There is a single independent basis state.
\begin{eqnarray}
\left( \chi_{12} \right)_{s=1} \otimes \left( \chi_{34} \right)_{s=1}
\end{eqnarray}
The symmetry property under the transpositions are given by
\begin{eqnarray}
\hspace{-0.8cm}(12)\left( \chi_{12} \right)_{s=1} \otimes \left( \chi_{34} \right)_{s=1} &=& (34)\left( \chi_{12} \right)_{s=1} \otimes \left( \chi_{34} \right)_{s=1}	\nonumber
\\
&=&\left( \chi_{12} \right)_{s=1} \otimes \left( \chi_{34} \right)_{s=1}.
\end{eqnarray}

In general, the color-spin space for $S=0$ is spanned by the following four color-spin bases.
\begin{eqnarray}
\phi_1 &=& \left( q_1 q_2 \right)^{6} \otimes \left( \bar{q}_3 \bar{q}_4 \right)^{\bar{6}} \left( \chi_{12} \right)_{s=1} \otimes \left( \chi_{34} \right)_{s=1}	\nonumber
\\
&\equiv& \left( q_1 q_2 \right)^{6}_{1} \otimes \left( \bar{q}_3 \bar{q}_4 \right)^{\bar{6}}_{1},	\nonumber
\\
\phi_2 &=& \left( q_1 q_2 \right)^{\bar{3}} \otimes \left( \bar{q}_3 \bar{q}_4 \right)^{3} \left( \chi_{12} \right)_{s=1} \otimes \left( \chi_{34} \right)_{s=1}	\nonumber
\\
&\equiv& \left( q_1 q_2 \right)^{\bar{3}}_{1} \otimes \left( \bar{q}_3 \bar{q}_4 \right)^{3}_{1},	\nonumber
\\
\phi_3 &=& \left( q_1 q_2 \right)^{6} \otimes \left( \bar{q}_3 \bar{q}_4 \right)^{\bar{6}} \left( \chi_{12} \right)_{s=0} \otimes \left( \chi_{34} \right)_{s=0}	\nonumber
\\
&\equiv& \left( q_1 q_2 \right)^{6}_{0} \otimes \left( \bar{q}_3 \bar{q}_4 \right)^{\bar{6}}_{0},	\nonumber
\\
\phi_4 &=& \left( q_1 q_2 \right)^{\bar{3}} \otimes \left( \bar{q}_3 \bar{q}_4 \right)^{3} \left( \chi_{12} \right)_{s=0} \otimes \left( \chi_{34} \right)_{s=0}	\nonumber
\\
&\equiv& \left( q_1 q_2 \right)^{\bar{3}}_{0} \otimes \left( \bar{q}_3 \bar{q}_4 \right)^{3}_{0}.
\end{eqnarray}
Similarly, there are six bases for $S=1$.
\begin{eqnarray}
\hspace{-0.6cm}
&\psi_1 = \left( q_1 q_2 \right)^{6}_{1} \otimes \left( \bar{q}_3 \bar{q}_4 \right)^{\bar{6}}_{1}
, \,\,
\psi_2 = \left( q_1 q_2 \right)^{\bar{3}}_{1} \otimes \left( \bar{q}_3 \bar{q}_4 \right)^{3}_{1} , &
\nonumber \\
\hspace{-0.6cm}
&\psi_3 = \left( q_1 q_2 \right)^{6}_{1} \otimes \left( \bar{q}_3 \bar{q}_4 \right)^{\bar{6}}_{0}
, \,\,
\psi_4 = \left( q_1 q_2 \right)^{\bar{3}}_{1} \otimes \left( \bar{q}_3 \bar{q}_4 \right)^{3}_{0} , &
\nonumber \\
\hspace{-0.6cm}
&\psi_5 = \left( q_1 q_2 \right)^{6}_{0} \otimes \left( \bar{q}_3 \bar{q}_4 \right)^{\bar{6}}_{1}
, \,\,
\psi_6 = \left( q_1 q_2 \right)^{\bar{3}}_{0} \otimes \left( \bar{q}_3 \bar{q}_4 \right)^{3}_{1}. &
\end{eqnarray}
The basis of the Hamiltonian is determined to satisfy the symmetry constraint due to the Pauli's principle, and constructed by combining the color-spin basis with the spatial part. The color-spin bases for each tetraquark state are listed in Table~\ref{result}.


\section{NUMERICAL RESULTS}

We substitute the wave function and perform a variational analysis to determine the ground state parameters in Eq.~(\ref{spatial-wave}).  The results are shown in Table~\ref{result}.   The binding energy is defined as 
\begin{eqnarray}
B_T & = & M_{Tetraquark} -M_{meson-1} -M_{meson-2},
\end{eqnarray}
where $M_{meson-1},M_{meson-2}$ are the masses of the two lowest  s-wave  states allowed in the decay of the tetraquark state.  For the spin 0 tetraquark states,  they are the two pseudo-scalar mesons, while for the spin-1 states, they are the pseudo-scalar and vector mesons.

As dicussed in Ref.~\cite{Park:2013fda}, the mixing terms appearing in the hyperfine potential part reduce the ground state energy of the $qq\bar{Q}\bar{Q}$ tetraquarks. As is shown in Table~\ref{result}, tetraquark states with $(I,S)=(0,1)$ are the most stable states.  Here the variational method is used to find the ground state from the lowest eigenvalue.

\subsection{Spatial size of the tetraquarks}

\begin{table}[t]

\caption{The contribution from each term in the Hamiltonian and the relative lengths between quarks in $ud\bar{c}\bar{c}$ with $(I,S)=(0,1)$, and in the lowest threshold mesons($\bar{D}^0 D^{*-}$).  Here, $V^C$= Coulomb+Linear interaction, and $(i,j)$ denotes the contribution from the $i$ and $j$ quark. The number is given as $i$=1, 2 for the light quarks, and 3, 4 for $\bar{c}$. The contributions are  in MeV unit.}

\centering

\begin{tabular}{ccccc}
\hline
\hline  & $(i,j)$ &  $ud\bar{c}\bar{c}$ \quad & \,\, 2-Meson \,\, & 	Difference	\quad	\\
\hline
Kinetic Energy							& 			&	1016.1 	&	880.4	&	135.7	\\
CS Interaction							&			&	-174.3	&	-73.6	&	-100.7	\\
\hline
										&	(1,2)	&	219.9	&		\\
										&	(1,3)	&	93.5	&	229.5 ($\bar{D}^0$)	\\
\multirow{3}{*}{$V^C$}				&	(1,4)	&	93.5	&		\\
										&	(2,3)	&	93.5	&		\\
										&	(2,4)	&	93.5	&	308.0 ($D^{*-}$)	\\
										&	(3,4)	&	15.6	&		\\
\cline{2-5}
										&Subtotal	&	609.5	&	537.5	&	72.0	\\
\hline
Total Contribution					&			&	1451.3	&	1344.3	&	107.0	\\
\hline
\hline
							 			&	(1,2)	&	0.67	&		\\
\multirow{3}{*}{Relative}			&	(1,3)	&	0.63	&	0.53 ($\bar{D}^0$)	\\
\multirow{3}{*}{Lengths}				&	(1,4)	&	0.63	&		\\
\multirow{3}{*}{(fm)}					&	(2,3)	&	0.63	&		\\
										&	(2,4)	&	0.63	&	0.58 ($D^{*-}$)	\\
										&	(3,4)	&	0.41	&		\\
\cline{2-5}
										&	Average&	0.60	&	0.56	&	0.04	\\
\hline 
\hline
\label{udccRel}
\end{tabular}
\end{table}

\begin{table}[t]

\caption{The contribution from each term in the Hamiltonian and the relative lengths between quarks in $ud\bar{b}\bar{b}$ with $(I,S)=(0,1)$, and in the lowest threshold mesons($B^+ B^{*0}$).  Here, $V^C$= Coulomb+Linear interaction, and $(i,j)$ denotes the contribution from the $i$ and $j$ quark. The number is given as $i$=1, 2 for the light quarks, and 3, 4 for $\bar{b}$. The contributions are expressed in MeV unit.}

\centering

\begin{tabular}{ccccc}
\hline
\hline  & $(i,j)$ &  $ud\bar{b}\bar{b}$ \quad & \,\, 2-Meson \,\, &		Difference	\quad	\\
\hline
Kinetic Energy							& 			&	997.2	&	836.6	&	160.6	\\
CS Interaction							&			&	-176.8	&	-26.4	&	-150.4	\\
\hline
										&	(1,2)	&	219.9	&		\\
										&	(1,3)	&	83.5	&	229.5 ($B^+$)	\\
\multirow{3}{*}{$V^C$}				&	(1,4)	&	83.5	&		\\
										&	(2,3)	&	83.5	&		\\
										&	(2,4)	&	83.5	&	266.6 ($B^{*0}$)	\\
										&	(3,4)	&	-187.6	&		\\
\cline{2-5}
										&Subtotal	&	366.3	&	496.1	&	-129.8	\\
\hline
Total Contribution					&			&	1186.7	&	1306.3	&	-119.6	\\
\hline
\hline
							 			&	(1,2)	&	0.67	&			\\
\multirow{3}{*}{Relative}			&	(1,3)	&	0.60	&	0.53 ($B^+$)	\\
\multirow{3}{*}{Lengths}				&	(1,4)	&	0.60	&			\\
\multirow{3}{*}{(fm)}					&	(2,3)	&	0.60	&			\\
										&	(2,4)	&	0.60	&	0.55 ($B^{*0}$)	\\
										&	(3,4)	&	0.25	&			\\
\cline{2-5}
										&	Average&	0.55	&	0.54	&	0.01	\\
\hline 
\hline
\label{udbbRel}
\end{tabular}
\end{table}

\begin{table}[t]

\caption{The contribution from each term in the Hamiltonian and the relative lengths between quarks in $us\bar{b}\bar{b}$ with $(I,S)=(1/2,1)$, and in the lowest thrshold mesons($B^0_s B^{*+}$). Here, $V^C$= Coulomb+Linear interaction, and $(i,j)$ denotes the contribution from the $i$ and $j$ quark. The number is given as $i$=1 for the light quark, 2 for $s$, and 3, 4 for $\bar{b}$. The contributions are expressed in MeV unit.}

\centering

\begin{tabular}{ccccc}
\hline
\hline  & $(i,j)$ & $us\bar{b}\bar{b}$ \quad & \,\,  2-Meson \,\, & 	Difference	\quad	\\
\hline
Kinetic Energy							& 			&	935.9	&	817.3	&	118.6	\\
CS Interaction							&			&	-118.2	&	-26.3	&	-91.9	\\
\hline
										&	(1,2)	&	171.3	&		\\
										&	(1,3)	&	56.3	&	266.6 ($B^{*+}$)	\\
\multirow{3}{*}{$V^C$}				&	(1,4)	&	56.3	&		\\
										&	(2,3)	&	77.5	&		\\
										&	(2,4)	&	77.5	&	21.1 ($B^0_s$)	\\
										&	(3,4)	&	-184.5	&		\\
\cline{2-5}
										&Subtotal	&	254.4	&	287.7	&	-33.3	\\
\hline
Total Contribution					&			&	1072.1	&	1078.7	&	-6.6	\\
\hline
\hline
							 			&	(1,2)	&	0.60	&			\\
\multirow{3}{*}{Relative}			&	(1,3)	&	0.52	&	0.55 ($B^{*+}$)	\\
\multirow{3}{*}{Lengths}				&	(1,4)	&	0.52	&			\\
\multirow{3}{*}{(fm)}					&	(2,3)	&	0.58	&			\\
										&	(2,4)	&	0.58	&	0.40 ($B^0_s$)	\\
										&	(3,4)	&	0.25	&			\\
\cline{2-5}
										&	Average&	0.51	&	0.48	&	0.03	\\
\hline 
\hline
\label{usbbRel}
\end{tabular}
\end{table}

\begin{table}[t]

\caption{The contribution from each term in the Hamiltonian and the relative lengths between quarks in $\Lambda^+_c$ and $\Xi^{++}_{cc}$. Here, $V^C$= Coulomb+Linear interaction,  and $(i,j)$ denotes the contribution from the $i$ and $j$ quark. For $\Lambda^+_c$, $i$=1, 2 for the light quark, 3 for $c$, while for $\Xi^{++}_{cc}$, the labelling is reversed.   }

\centering

\begin{tabular}{cccc}
\hline
\hline  & $(i,j)$ & $\Lambda^+_c$ \quad	& \quad $\Xi^{++}_{cc}$ \quad \\
\hline
Kinetic Energy (MeV)					& 			&	797.7		&	658.3		\\
CS Interaction (MeV)					&			&	-181.8		&	-51.0		\\
\hline
										&	(1,2)	&	219.9		&	5.6	\\
\multirow{2}{*}{$V^C$ (MeV)}			&	(1,3)	&	176.0		&	156.0	\\
										&	(2,3)	&	176.0		&	156.0	\\
\cline{2-4}
										&Subtotal	&	571.9		&	317.6	\\
\hline
Total Contribution					&			&	1187.8		&	924.9	\\
\hline
\hline
\multirow{2}{*}{Relative}			&	(1,2)	&	0.67		&	0.40	\\
\multirow{2}{*}{Lengths}				&	(1,3)	&	0.61		&	0.58	\\
\multirow{2}{*}{(fm)}					&	(2,3)	&	0.61		&	0.58	\\
\cline{2-4}
										&	Average&	0.63		&	0.52	\\
\hline 
\hline
\label{lambdaXiRel}
\end{tabular}
\end{table}

To understand the composition of the total energy of the tetraquark states in comparison to the lowest threshold decaying hadrons, it is important to understand the spatial size of the systems. These determine the  magnitude of the various kinetic energies and the potential energies between quarks. We analyze  the contributions in the lowest energy color spin state for each quantum number.

Let us first discuss the kinetic energy part.  As an example, let us discuss the case for $ud \bar{c} \bar{c}$ as given in Table~\ref{udccRel}.  The kinetic energy of the tetraquark state obtained as 1016.1 MeV can be understood as the sum of three internal kinetic energies: kinetic energies of the $u-\bar{c}$, $d -\bar{c}$, and the $(u\bar{c})-(d \bar{c})$ pairs. The sum of the internal kinetic energies of the $\bar{D} D^*$ states is coming from  $u-\bar{c}$, $d -\bar{c}$.  Therefore, the tetraquark system has an extra kinetic energy needed to bring the $\bar{D} D^*$ system into a compact configuration. The actual kinetic energy of the $u-\bar{c}$ ($d -\bar{c}$) in the tetraquark system is smaller than that inside the $\bar{D} (D^*)$ system. This is so because as can be seen in Table~\ref{udccRel} the size of this pair is larger in the tetraquark system than in the meson: the distance $(1,3)$ is 0.63 fm in the tetrquark system while it is 0.53 fm in $\bar{D}$.  Still, the additional kinetic energy makes the sum of the kinetic energies in the tetraquark system to be larger than the total kinetic energy of the threshold mesons. This is one of the reasons why we need large attraction to bind a multiquark system into a compact configuration.  Such mechanism is also present in other heavier tetraquark systems.
In general, the additional kinetic energy is of the following form.
\begin{eqnarray}
K_{add} = \frac{3 {\mathbf p}_{\mathbf x}^2}{2 \mu},
\end{eqnarray}
where  ${\mathbf p}_{\mathbf x}$ is the relative momentum of the mesons, which is inversely proportional to the size of the pair of heavy quarks and thus increases as the heavy quark mass increases. $\mu$ is the reduced mass of the two meson masses, which also increases as the heavy quark mass increases. Hence, the additional kinetic energy seems to be present for all quark masses.

Let us now turn our discussion to the potential parts. 
In most of the tetraquark configurations, the contributions to the bound state from the parts of $V^C$ and kinetic energy are repulsive so that the contribution from the color spin interaction becomes important. However, the contributions from $V^C$ for the two bound states in Tables~\ref{udbbRel}, \ref{usbbRel} are attractive. 
Thus, for a better understanding of the binding mechanism, we concentrate on those attractive contributions from $V^C$. 
Comparing  Table~\ref{udccRel} and Table~\ref{udbbRel}, one notes that the $V^C$ between the heavy quarks is much more attractive in $ud\bar{b}\bar{b}$ than in $ud\bar{c}\bar{c}$.  This is an important result which was used as an input in Ref.~\cite{Karliner:2017qjm}.
The reason for the difference in binding comes from the smaller relative length between the pair $\bar{b}-\bar{b}$ which is 0.61 times  that of  $\bar{c}-\bar{c}$. 
 The length between $\bar{b} - \bar{b}$ is also very short compared to the quark-antiquark distance inside the two threshold mesons: the distance between $\bar{b}-\bar{b}$ pair is 0.25 fm which is 0.46 times the average value of 0.54 fm in the two threshold mesons.  On the other hand, for the $ud\bar{c}\bar{c}$ case, the distance between $\bar{c}-\bar{c}$ is smaller than the corresponding two threshold mesons by a lesser degree. 

Now we look at $us\bar{b}\bar{b}$ given in Table~\ref{usbbRel}, which is weakly bound compared to $ud\bar{b}\bar{b}$. In this case, $d$ quark in $ud\bar{b}\bar{b}$ is replaced by the heavier $s$ quark so that one needs to consider the binding effect from the pair $s - \bar{b}$ as already discussed in Ref.~\cite{Karliner:2014gca}. Comparing Table~\ref{usbbRel} to Table~\ref{udbbRel}, one finds that $V^C$ and most of  the relative lengths in $us\bar{b}\bar{b}$ are reduced. At the same time, the interaction $V^C$ in $B_s$ which is the lowest threshold meson in the $us\bar{b}\bar{b}$ decay  becomes much more attractive than that in $B$ so  that the total contribution from $V^C$ is not as attractive as in $ud\bar{b}\bar{b}$. In addition, the attraction from the color spin part also becomes smaller than that in $ud\bar{b}\bar{b}$. On the other hand, the contribution from the kinetic energy part is less repulsive than in $ud\bar{b}\bar{b}$ because the additional kinetic energy becomes smaller, due to the increase in the reduced mass,  but the sum of the kinetic energies in the threshold mesons are similar in both cases.
 However, this is not enough to make up the above repulsive changes in both the color spin and $V^C$ parts, leading to the barely bound $us\bar{b}\bar{b}$ state.

For the values of the total contributions in Tables~\ref{udccRel}-\ref{usbbRel}, we need to consider an additional binding effect coming from the mixing terms between different color spin states appearing in the hyperfine potential part of the Hamiltonian. Considering the additional effect, one can get the binding energies for each tetraquark state given in Table~\ref{result}.

\subsection{Comparison with a simple quark model}

As emphasized in the introduction, it is essential to perform a full constituent quark model analysis to calculate the mass and binding energy of a multiquark configuration. For that purpose, let us compare the composition of the mass of three states $\Xi_{cc}$, $T_{QQ}$, $\Lambda_c$ calculated from our model and from a simple constituent quark model~\cite{Karliner:2014gca}. 
In a simple quark model, as discussed in Ref.~\cite{Karliner:2014gca}, the mass of a hadron is typically composed of the sum of effective constituent quark masses, the hyperfine interaction, and a possible binding energy for heavier quarks.    We want to identify the origin of the effective constituent quark mass and the binding energy used in the simple model from our full model calculation, and then investigate whether it is sensible to extrapolate these concepts to higher multiquark configurations.

In the simple constituent quark model, the mass of $\Lambda_c$, $\Xi_{cc}$ can be obtained from the following formula\cite{Karliner:2014gca}.
\begin{eqnarray}
M_{\Lambda_c} & = & 2m_q^b+m_c^b -\frac{3a}{(m^b_q)^2}, \nonumber \\
M_{\Xi_{cc}} & = & 2m_c^b+B(cc)+m_q^b+ \frac{a_{cc}}{(m^b_c)^2} -\frac{4a}{m^b_q m^b_c}, \label{simple}
\end{eqnarray}
where $m^b_{c, q}$ are the constituent quark masses for the charm and light quark inside a baryon, $B(cc)$ the binding between the charm quarks, and $a$'s  multiplicative constants for the color-spin interaction. Treating $B(cc)$ as part of the two charm quark system, one can dive the energy into the charm quark, light quark and color-spin interaction parts. 

 Let us now see how to understand the three parts in our full constituent quark model.   As can be seen from Eq.~(\ref{hamiltonian}), the model is based on two body interaction and cannot in general be divided in the three pieces.  Nevertheless, one can calculate the total mass according to Eq.~(\ref{hamiltonian}) and group the individual terms into `$q$-quark', `$c$-quark' and `color-spin(CS)' parts as given in Tables \ref{OurXicc}-\ref{OurTcc}.

First, few explanations of the division of terms in Tables \ref{OurXicc} and \ref{OurLambdaC} are in order.  The constant $-D$ term appearing in Eq.~(\ref{potential1}) is divided into each quark by multiplying a factor of 1/2.  This is so because when the total hadron is a color singlet, the total color factors contribute equally for all quarks involved.  For the kinetic terms, when they involve the quark pairs, it is included in the corresponding quark pairs.  For the relative kinetic energy involving ${\mathbf p}_{\boldsymbol{\lambda}}$, it is divided according to their relative contribution depending on the mass of either the quark pair or the single quark.   

Let us now compare the values in the third column and the fifth column (given in Eq.~(\ref{simple})) in Tables~\ref{OurXicc} and \ref{OurLambdaC}. We are not interested in the detailed numbers, but the systematics involved in the comparison.   The following are important points to note.

\begin{enumerate}
\item  $V^C(1,2)$ in $\Xi_{cc}$ is much smaller than that in $\Lambda_c$. The heavy quark pair is much more compact than the light quark pairs and thus feels more attraction.

\begin{widetext}

\begin{table}

\caption{Division of the $\Xi_{cc}^{++}$ mass from our work.  $(i,j)$ denotes the contribution from the $i,j$ quarks, where $i$= 1, 2 labels  $c$, and 3  the light quark. Here, $V^c$= Coulomb+Linear interaction, $m'_1=m_c, \,\, m'_2=\frac{3 m_q m_c}{m_q + 2m_c}, \,\, {\mathbf p}_{\boldsymbol{\sigma}} = m'_1 \dot{\boldsymbol{\sigma}}, \,\,
{\mathbf p}_{\boldsymbol{\lambda}} = m'_2 \dot{\boldsymbol{\lambda}},	$ and $ \boldsymbol{\sigma} = \frac{1}{\sqrt{2}}({\mathbf r}_1 - {\mathbf r}_2), \,\, \boldsymbol{\lambda} = \frac{1}{\sqrt{6}}({\mathbf r}_1+{\mathbf r}_2-2{\mathbf r}_3) $. All the values are expressed in MeV unit.}	
\centering
\begin{tabular}{c|cc|cc}
\hline
\multirow{2}{*}{Overall}	 & \multicolumn{2}{c|}{Present Work} & \multicolumn{2}{c}{Ref.~\cite{Karliner:2014gca}}	\\
\cline{2-3} \cline{4-5}
			& \,\,\, Contribution \,\,\, & \,\,\, Value \,\,\, & \,\,\, Contribution \,\,\, & \,\,\, Value \,\,\, \\
\hline 
$c$-quark	&								$2m_c$					&	3836.0	&	$2m_c^b$ &  3421.0	\\
			&	$\frac{{\mathbf p}_{\boldsymbol{\sigma}}^2}{2m'_1}$			&	243.6	&			\\
&$\frac{m_q}{m_q+2m_c}\frac{{\mathbf p}_{\boldsymbol{\lambda}}^2}{2m'_2}$		&	32.5	&		\\
			&								$V^C(1,2)$					&	5.6		& $B(cc)$ &	-129.0	\\
			&	$\frac{1}{2}\left[ V^C(1,3) + V^C(2,3) \right]$	&	156.0	&			\\	
			&								$-D$						&	-983.0 	&			\\
Subtotal		&														&	3290.7	& & 	3292.0	\\ \hline
$q$-quark		&								$m_q$					&	326.0	&	$m^b_q$	&	363.0		\\
&$\frac{2m_c}{m_q+2m_c}\frac{{\mathbf p}_{\boldsymbol{\lambda}}^2}{2m'_2}$		&	382.2	&	\\
			&	$\frac{1}{2}\left[ V^C(1,3) + V^C(2,3) \right]$	& 	156.0	&			\\
			&							$-\frac{1}{2} D$				&	-491.5	&			\\
Subtotal		&  														&	372.7	&	&	363.0		\\	 \hline
CS Interaction	&						$-4a/(m_q m_c)$					&	-58.8	&	$-4a/(m^b_q m^b_c)$	&	-42.4	\\
			&						$a_{cc}/(m_c)^2$					&	7.8		&	$a_{cc}/(m^b_c)^2$	&14.2	\\
Subtotal	&  															&	-51.0		&	&	-28.2	\\      \hline 
Total		&															&	3612.4	&	&	3626.8	\\
\hline
\multicolumn{2}{c}{}
\label{OurXicc}
\end{tabular}
\end{table}

\begin{table}

\caption{Division of the $\Lambda^+_c$ mass from our work. Notations are similar as in Table \ref{OurXicc}.  $i$= 1, 2 labels the light quarks, and 3 $c$.  $ m'_1=m_q, \,\, m'_2=\frac{3 m_q m_c}{2 m_q + m_c}, \,\, {\mathbf p}_{\boldsymbol{\sigma}} = m'_1 \dot{\boldsymbol{\sigma}}, \,\, {\mathbf p}_{\boldsymbol{\lambda}} = m'_2 \dot{\boldsymbol{\lambda}}$ and $ \boldsymbol{\sigma} = \frac{1}{\sqrt{2}}({\mathbf r}_1 - {\mathbf r}_2), \,\, \boldsymbol{\lambda} = \frac{1}{\sqrt{6}}({\mathbf r}_1+{\mathbf r}_2-2{\mathbf r}_3) $. All the values are expressed in MeV unit.}
\centering
\begin{tabular}{c|cc|cc}
\hline
\multirow{2}{*}{Overall}	 & \multicolumn{2}{c|}{Present Work} & \multicolumn{2}{c}{Ref.~\cite{Karliner:2014gca}}	\\
\cline{2-3} \cline{4-5}

		 					 & \quad Contribution \quad & \,\, Value \,\, & \quad Contribution \quad & \,\, Value \,\,	\\
\hline  
$q$-quark	&							$2m_q $								&	652.0	&	$2m^b_q$	&	726.0		\\
			&	$\frac{{\mathbf p}_{\boldsymbol{\sigma}}^2}{2m'_1}$	&	501.7	&			\\
&$\frac{m_c}{2m_q+m_c}\frac{{\mathbf p}_{\boldsymbol{\lambda}}^2}{2m'_2}$&	221.0	&			\\
			&						$V^C(1,2)$								&	219.9 	&			\\
			&	$\frac{1}{2} \left[ V^C(1,3) + V^C(2,3) \right]$		&	176.0	&			\\
			&							$-D$								&	-983.0	&			\\
Subtotal	&																&	787.6	&	&	726.0		\\ \hline
$c$-quark	&							$m_c$								&	1918.0 &	$m^b_c$	&	1710.5	\\
&$\frac{2m_q}{2m_q+m_c}\frac{{\mathbf p}_{\boldsymbol{\lambda}}^2}{2m'_2}$&	75.1&		\\
			&	$\frac{1}{2} \left[ V^C(1,3) + V^C(2,3) \right]$		&	176.0	&			\\
			&						$-\frac{1}{2} D$						&	-491.5	&			\\
Subtotal	&																&	1677.6 &	&	1710.5	\\	 \hline
CS Interaction&					$-3a/(m_q)^2$							&	-181.8	&	$-3a/(m^b_q)^2$	&	-150.0	\\     \hline 
Total		&																&	2283.4	&	&	2286.5	\\
\hline
\multicolumn{2}{c}{}
\label{OurLambdaC}
\end{tabular}
\end{table}

\begin{table}[t]

\caption{Division of the $ud\bar{c}\bar{c}$ mass from our work.  $(i,j)$ denotes the $i$ and $j$ quarks, where $i$= 1, 2 labels the light quarks, and 3, 4 are for $\bar{c}$. $ m'_1=m_q, \,\, m'_2=m_c, \,\, m'_3=\frac{2 m_q m_c}{m_q + m_c}, {\mathbf p}_{\boldsymbol{\sigma}} = m'_1 \dot{\boldsymbol{\sigma}}, \,\,
{\mathbf p}_{\boldsymbol{\sigma}'} = m'_2 \dot{\boldsymbol{\sigma}'}, \,\,
{\mathbf p}_{\boldsymbol{\lambda}} = m'_3 \dot{\boldsymbol{\lambda}}$ and $
 \boldsymbol{\sigma} = \frac{1}{\sqrt{2}}({\mathbf r}_1 - {\mathbf r}_2), \,\, \boldsymbol{\sigma}' = \frac{1}{\sqrt{2}}({\mathbf r}_3 - {\mathbf r}_4) , \,\, \boldsymbol{\lambda} = \frac{1}{2} \left({\mathbf r}_1 + {\mathbf r}_2 - {\mathbf r}_3 - {\mathbf r}_4 \right) $. The estimate in the fourth column is obtained by taking the $c$-quark pair value of ($\Xi_{cc}$) and $q$-quark pair value of 
 $\Lambda_c$ from our model calculation. All the values are expressed in MeV unit.}	
\centering
\begin{tabular}{c|ccc|cc}
\hline
\multirow{2}{*}{Overall}	 & \multicolumn{3}{c|}{Present Work} & \multicolumn{2}{c}{Ref.~\cite{Karliner:2017qjm}}	\\
\cline{2-4} \cline{5-6}

		 					 & \quad Contribution \quad & \,\, Value \,\, & \,\, Estimate \,\, & \quad Contribution \quad & \,\, Value \,\,	\\
\hline 
$c$-quark	&					$2m_c$							&	3836.0	&	3836.0	& $2m_c^b$ & 	3421.0	\\
			&	$\frac{{\mathbf p}_{\boldsymbol{\sigma}'}^2}{2m'_2}$	&	231.4	&	243.6	&  		&	\\
			
			&$\frac{m_q}{m_c+m_q} \frac{{\mathbf p}_{\boldsymbol{\lambda}}^2}{2  m'_3}$&	41.1	&	32.5	&  	&	\\ 
			
			&							$V^C(3,4)$							&	15.6	&	5.6		& $B(cc)$ & 	-129.0	\\
			
&$\frac{1}{2} \left[ V^C(1,3) + V^C(1,4) + V^C(2,3) + V^C(2,4) \right]$	&	187.0	&	156.0	&	&	\\   
			
			&								$-D$							&	-983.0	&	-983.0	&	&	\\
			
Subtotal	&																&	3328.1	&	3290.7	&   & 3292.0	\\
\hline

$q$-quark	&					$2m_q$								&	652.0	&	652.0	&	$2m^b_q$	&	726.0	\\   

			&	$\frac{{\mathbf p}_{\boldsymbol{\sigma}}^2}{2m'_1}$	&	501.7	&	501.7	&	&	\\  
			
			&$\frac{m_c}{m_c+m_q} \frac{{\mathbf p}_{\boldsymbol{\lambda}}^2}{2  m'_3}$&	241.9	&	221.0	&		&	\\
			
			&						$ V^C(1,2)$							&	219.9	&	219.9	&	&	\\
			
&$\frac{1}{2} \left[ V^C(1,3) + V^C(1,4) + V^C(2,3) + V^C(2,4) \right]$&	187.0	&	176.0	&	&	\\   
			
			&								$-D$							&	-983.0	&	-983.0	&	&	\\
			
Subtotal	&																&	819.5	&	787.6	& 	&	726.0 \\
\hline

CS Interaction	&				$a_{cc}/(m_c)^2$						&	7.5		&	7.8		&	$a_{cc}/(m^b_c)^2$ &	14.2 \\

					&				$-3a/(m_q)^2$						&	-181.8	&	-181.8	& 	$-3a/(m^b_q)^2$&	-150.0 \\         
					   
Subtotal	&																&	-174.3	&	-174.0	& 	&	-135.8 \\
\hline

\multirow{2}{*}{Total}	&		\multirow{2}{*}{}		&	\multirow{2}{*}{3973.3}	& \multirow{2}{*}{3904.3}	& \multirow{2}{*}{}	&	\multirow{2}{*}{3882.2} \\

	& & & & & \\
\hline
\multicolumn{2}{c}{}
\label{OurTcc}
\end{tabular}
\end{table}

\end{widetext}

This can be understood as the origin of attraction for the heavy quark pair as given by -129 MeV in the simple quark model of Ref.~\cite{Karliner:2017qjm}.

\item  Our model's subtotal value of the $q$-quark ($c$-quark) in $\Xi_{cc}$ ($\Lambda_c$) can be thought of the constituent quark in the simple quark model.  Multiplying our value for the $q$-quark ($c$-quark) value in $\Xi_{cc}$ ($\Lambda_c$)  by a factor 2 gives approximately the value of the $q$-quark ($c$-quark) pair in $\Lambda_c$ ($\Xi_{cc}$) of our model with slight repulsion (attraction), pointing to the need to introduce an additional attraction for the heavy pair with respect to the light quark pair in a simple constituent quark model.

\end{enumerate}

In conclusion, as can be seen by comparing the third and fifth column of Table \ref{OurXicc} and \ref{OurLambdaC}, the constituent quark masses and the binding energy as needed in Eq.~(\ref{simple}) should be the sum of the quark mass, the relevant kinetic term, and all the relevant interaction terms in the full model, which indeed seems to approximately reproduce the simple constituent quark mass value.  

Now let us see what happens when we try to build up a similar table for $T_{cc}$.
According to the simple constituent quark model, the mass is given as\cite{Karliner:2017qjm}
\begin{eqnarray}
M_{T_{cc}} & = &  2m_c^b+B(cc)+2m_q^b+ \frac{a_{cc}}{(m^b_c)^2} -\frac{3a}{(m^b_q)^2}. \label{simple2}
\end{eqnarray}
Apart from the color-spin interaction part, it is the sum of the subtotal mass of $(cc)$ pair in $\Xi_{cc}$ and $(qq)$ pair in $\Lambda_c$. This is so because the color spin state of $(cc)$ pair in $\Xi_{cc}$   is the same as that of the charge conjugated $(\bar{c} \bar{c})$ pair in the lowest energy component of $ud\bar{c}\bar{c}$ state.  Similarly, the color spin state of $(ud)$ pair  is the same as that of the light quark pair in the lowest energy component of $ud\bar{c}\bar{c}$ state. Table \ref{OurTcc}  shows the numbers in our model calculations. The estimate in the  fourth column is obtained by taking such prescription in our model.  
As can be seen in the table, the estimate is much closer to the simple quark model result of Ref.~\cite{Karliner:2017qjm}.  On the  other hand, the our value from the full model calculation in column three is systematically larger, except for $\frac{{\mathbf p}_{\boldsymbol\sigma '}^2}{2m'_2}$, than the other estimates. 
The change comes from the slight differences in the potentials. 
 The dominant change comes in from the kinetic energy of the  relative momentum ${\mathbf p}_{\boldsymbol{\lambda}}$.  The magnitude of this relative kinetic energy and its contribution to each pair depends on the masses of each pair and cannot be extrapolated from an estimate obtained within a hadron of different quark numbers.  This is related to the changes of the effective constituent quark mass in a simple quark model, which indeed shows the need for different values for the quark masses depending on whether they are inside a meson or a baryon.  
Ref.~\cite{Karliner:2014gca} finds that the constituent quark masses to be $m_q^m=310$ MeV, $m_s^m=483$ MeV, $m_c^m=1663.3$ MeV to fit the meson spectrum,  while they are $m_q^b=363$ MeV, $m_s^b=538$ MeV, $m_c^b=1710.5$ MeV to fit the baryon spectrum.  The trend of needing larger masses when it is inside configurations with larger constituent seems also to be true when one goes to the tetraquark configuration as can be seen in our full model calculation shown in Table \ref{OurTcc}.  Our estimates are systematically larger than the simple quark model estimates.

\section{summary}

The question for the possbile existence of  stable $qq\bar{Q}\bar{Q}$ tetraquarks has a long history.  With the recent discovery of the doubly charmed baryon $\Xi_{cc}$, it now became possible to quantitatively calculate the binding energy of the stable $ud\bar{b}\bar{b}$. Indeed, based on a simplified constituent quark model, the authors in Ref.~\cite{Karliner:2017qjm} have recently shown that the  $bb\bar{u}\bar{d}$ tetraquark state is deeply bound, the  $cc\bar{u}\bar{d}$ is above the threshold and $bc\bar{u}\bar{d}$ is slightly below the threshold. 
The simplified model was based on approximating the tetraquark mass to be the sum of the constituent quark masses and the hyperfine interaction.  In this work, we have performed a full constituent quark model calculation 
 with specific potentials, as given in Eqs.~(\ref{hamiltonian})-(\ref{potential4}), and  using the variational method with one gaussian, to fit the mass of $\Xi_{cc}$ and other related hadrons to study the stability of these tetraquark states.   
By comparing the results of the two approaches for the masses of $\Lambda_c$ and $\Xi_{cc}$, we have shown that the constituent quark mass in the simplified model can be identified as the sum of the quark mass, the relevant kinetic term and all the relevant interaction terms in our full model calculation. 
We have then found that the tetraquark masses calculated in our model are systematically larger than those estimated in the simple model calculations. 
Comparing our full model calculations for the tetrqaurk states to those from the simplified model approach, we also found that when using a simplified quark model, it is necessary to introduce a slightly larger constituent quark mass in the tetraquark configuration than in the baryon.  This is consistent with the trend where one needs a larger constituent quark mass in the baryon than in the meson. 
 
We found that the $ud\bar{b}\bar{b}$ tetraquark state is bound by 120.56 MeV and that the  $us\bar{b}\bar{b}$ slightly bound by 7.3 MeV, while all other tetraquark states are above threshold.  
Nevertheless, it is still possible that the tetraquark state $T_{cc}(ud\bar{c}\bar{c})$ is a resonance state.  Furthermore, all the tetraquark states discussed are flavor exotic, and it would be extremely interesting to confirm the existence of these states experimentally.

\section*{Acknowledgments}
This work was supported by the Korea National Research
Foundation under the grant number 2016R1D1A1B03930089 and 2017R1D1A1B03028419(NRF).

\end{document}